\documentclass[12pt]{article}\usepackage{EntDistil}

\begin{document}

\title{Entanglement of Distillation\\
and Conditional Mutual Information}

\author{Robert R. Tucci\\
        P.O. Box 226\\
        Bedford,  MA   01730\\
        tucci@ar-tiste.com}

\date{ \today}

\maketitle

\vskip2cm
\section*{Abstract}
In previous papers, we expressed the Entanglement of Formation
in terms of Conditional Mutual Information (CMI). In this brief paper,
we express the Entanglement of Distillation in terms of CMI.

\newpage
\section{Introduction}
Entanglement of Formation ($E_F$)
and Entanglement of Distillation ($E_D$)
were invented by Bennett et al in Ref.\cite{Ben} and
satellite papers. In a  series of previous papers\cite{Tucci-series},
we showed how to
express $E_F$ in terms  of
Conditional Mutual Information (CMI), but we said nothing about
$E_D$. In this brief letter, we will show how to express
$E_D$ in terms of CMI.
Recently, other researchers have
expressed some of  their entanglement ideas in terms of unconditional
mutual information. See, for example, Ref.\cite{IBM}.

Two reasons why CMI is useful for quantifying entanglement
are the following. First, entanglement is
 an ``exclusively quantum" effect. CMI satisfies this requirement. It
vanishes in the classical regime, but not in the quantum regime,
for a fiducial experiment.
Second, entanglement is associated with a correlation
between two events $\rva$ and $\rvb$. But there must be
something to distinguish entanglement correlations from
classical correlations.
CMI satisfies this requirement too. It measures more than just
the correlation of $\rva$ and $\rvb$.
Those two events are assumed to have a common ancestor event
(or cause, or antecedent) in their past, call  it $\rvlam$, and
we condition on that common ancestor. (See Fig.\ref{fig:ent-form})
For example, in Bohm's version of the EPR
experiment, $\rvlam$ might correspond to the
event of a spin-zero particle breaking up into
two spin-half particles with opposite spins.

We will try to make this paper as
self contained as we can for such a short document.
If the reader has any questions concerning notation or definitions,
we refer him to Ref.\cite{Tucci-Rev}---a
much longer, tutorial paper
that uses the same notation as this paper.

We will represent random variables by underlined letters.
$S_\rva$ will be the set of all possible values that $\rva$
can assume, and $N_\rva$ will be the number of elements in $S_\rva$.
$S_{\rva, \rvb}$ will represent
the Cartesian product of sets $S_\rva$ and $S_\rvb$.
In the quantum case, $\hil_\rva$ will represent a Hilbert space
of dimension $N_\rva$. $\hil_{\rva, \rvb}$ will represent the tensor
product of $\hil_\rva$ and $\hil_\rvb$. Red indices
should be summed over (e.g. $a_{\suma{i}} b_{\suma{i}} = \sum_{i}a_i b_i$).
$\pd(S_\rva)$ will denote the set of all probability distributions
on $S_\rva$,
$P(a)\geq 0 $ such that $\sum_{a\in S_\rva} P(a) = 1$.
$\dm(\hil_\rva)$
will denote the set of all density
matrices acting on $\hil_\rva$.

As usual\cite{Cover},
for any three random variables, $\rva, \rvb, \rvlam$,
we define the {\it mutual information}(MI) by

\beq
H(\rva: \rvb) =
H(\rva) + H(\rvb) - H(\rva, \rvb)
\;,
\eeq
and the {\it conditional mutual information}(CMI) by

\beqa
H(\rva : \rvb | \rvlam)
&=& H(\rva|\rvlam)
+ H(\rvb | \rvlam)
- H(\rva, \rvb | \rvlam)\nonumber\\
&=&
H(\rvlam)
-H(\rva, \rvlam)
-H(\rvb, \rvlam)
+H(\rva, \rvb, \rvlam)
\;.
\eeqa
Since $H(\rva|\rvlam)\leq H(\rva)$, one might be tempted
to assume that also $H(\rva : \rvb | \rvlam)\leq H(\rva : \rvb )$,
but this is not generally true. One can construct examples for
which CMI is greater or smaller than MI, a fact well known
since the early days of
Classical Information Theory\cite{McGill}, \cite{cmi-versus-mi}.

One can define analogous quantities for Quantum Physics.
Suppose $\rho_{\rva, \rvb, \rvlam}\in \dm(\hil_{\rva, \rvb, \rvlam})$,
with partial  traces
$\rho_\rva \in \dm(\hil_\rva)$, etc. Then we define

\beq
S(\rva:\rvb) =
S(\rho_{\rva})
+S(\rho_{\rvb})
-S(\rho_{\rva, \rvb})
\;,
\eeq
and

\beq
S(\rva:\rvb|\rvlam) =
S(\rho_\rvlam)
-S(\rho_{\rva, \rvlam})
-S(\rho_{\rvb, \rvlam})
+S(\rho_{\rva, \rvb, \rvlam})
\;.
\eeq

\section{Entanglement of Formation}
Before racing off at full speed, let's warm up with a brief review
of the CMI definition of $E_F$.

\begin{figure}[h]
    \begin{center}
    \epsfig{file=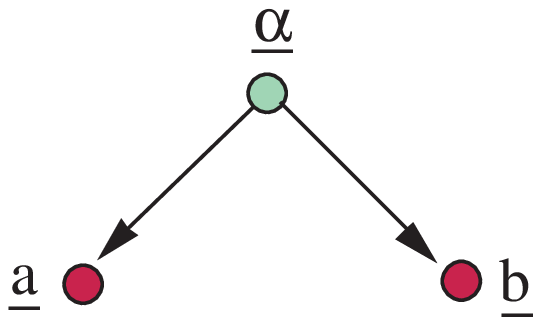, height=2.0in}
    \caption{Classical Bayesian Net that motivates
    the definition of $E_F$.}
    \label{fig:ent-form}
    \end{center}
\end{figure}

Consider the Classical Bayesian Net shown in Fig.\ref{fig:ent-form}.
It represents a probability distribution of the form:
\beq
P(a, b, \alpha) = P(a| \alpha) P(b | \alpha) P(\alpha)
\;.
\label{eq:prob-ent-form}\eeq
One can easily check that for this probability distribution,
$H(\rva: \rvb |\rvalp)$ is identically zero.

In the classical case, we define $E_F$ by
\beq
E_F(P_{\rva,\rvb}) = \min_{P_{\rva, \rvb, \rvalp}\in K} H(\rva: \rvb |\rvalp)
\;,
\label{eq:clas-ent-form}\eeq
where $K$ is the set of all
probability distributions $P_{\rva, \rvb, \rvalp}$
with a
fixed marginal $P_{\rva, \rvb}$.
Thus, $E_F$ is a function of $P_{\rva, \rvb}$.
If $K$ contains a
$P_{\rva, \rvb, \rvalp}$ of the
form given by Eq.(\ref{eq:prob-ent-form}),
then $E_F=0$. This is always true if $K$
is defined to contain all probability distributions
with arbitrary positive values of $N_\rvalp$.
But it may not be true if
$K$ contains only probability distributions with a
fixed $N_\rvalp$ value.
 The fact that the
right hand side of Eq.(\ref{eq:clas-ent-form})
vanishes
in the classical case (if $K$ includes all $N_\rvalp$ values)
 is an important motivation
for defining $E_F$ this way. We want
a measure of entanglement that
is exclusively quantum.

In the quantum case, suppose
$\{w_\alpha: \alpha=1, \ldots, N_\rvalp\}$ is a
probability distribution for $\rvalp$,
and $\{\ket{\alpha} : \alpha=1, \ldots, N_\rvalp\}$ is an orthonormal
basis for $\hil_\rvalp$.
For all $\alpha$, suppose $\rho^\alpha_\rva\in \dm(\hil_\rva)$,
and $\rho^\alpha_\rvb\in \dm(\hil_\rvb)$.
Consider a ``separable" density matrix $\rho_{\rva, \rvb, \rvalp}$
of the form

\beq
\rho_{\rva, \rvb, \rvalp}
=
\sum_\alpha
\rho_\rva^\alpha \rho_\rvb^\alpha
w_\alpha
\ket{\alpha}\bra{\alpha}
\;.
\label{eq:rho-ent-form}\eeq
One can easily check that for this density matrix, $S(\rva: \rvb | \rvalp)=0$.

In the quantum case, we define $E_F$ by

\beq
E_F(\rho_{\rva, \rvb}) = \min_{\rho_{\rva, \rvb, \rvalp}\in K}S(\rva: \rvb | \rvalp)
\;,
\label{eq:quan-ent-form}\eeq
where $K$ equals the set $K_0 =\{ \rho_{\rva, \rvb, \rvalp}
\in \dm(\hil_{\rva, \rvb, \rvalp}):$ arbitrary $N_\rvalp$,
fixed marginal
 $\rho_{\rva, \rvb}\}$.
 Thus, $E_F$ is a function of $\rho_{\rva, \rvb}$.
If $K$ contains a $\rho_{\rva, \rvb, \rvalp}$
of the form given by
Eq.(\ref{eq:rho-ent-form}),
then $E_F$ is zero. The quantum $E_F$ can be nonzero
even if $K$ contains all density matrices with arbitrary $N_\rvalp$
values.

In Eq.(\ref{eq:quan-ent-form}), we could set
$K= K_2$, where $K_2$ is the subset of $K_0$
which restricts $\rho_{\rva, \rvb, \rvalp}$
to be of the form

\beq
\rho_{\rva, \rvb, \rvalp}
=
\sum_\alpha
\ket{\psi^\alpha_{\rva,\rvb}}
\bra{\psi^\alpha_{\rva,\rvb}}
w_\alpha
\ket{\alpha}\bra{\alpha}
\;,
\eeq
where $\ket{\psi^\alpha_{\rva,\rvb}}\in \hil_{\rva,\rvb}$.
One can
show that Eq.(\ref{eq:quan-ent-form}) with $K= K_2$ is
identical (up to a factor of 2)
to the definition of $E_F$ originally given by
Bennett et al in Ref.\cite{Ben}.
Other possible $K$ choices come to mind.
For example, one could set $K$ equal to $K_1$,
where $K_1$ is that subset of $K_0$ which restricts $\rho_{\rva, \rvb, \rvalp}$
to be of the form

\beq
\rho_{\rva, \rvb, \rvalp}
=
\sum_\alpha
\rho^\alpha_{\rva,\rvb}
w_\alpha
\ket{\alpha}\bra{\alpha}
\;,
\eeq
where $\rho^\alpha_{\rva,\rvb}\in \dm(\hil_{\rva,\rvb})$
need not be pure.
$K_0, K_1, K_2$ represent different degrees of information about
how $\rho_{\rva, \rvb, \rvalp}$ was created. $K_0$ represents
total ignorance.

\section{Classical Distillation}
In this section, we will define a classical
$E_D$. In the next section, we will
find a quantum counterpart for it.

\begin{figure}[h]
    \begin{center}
    \epsfig{file=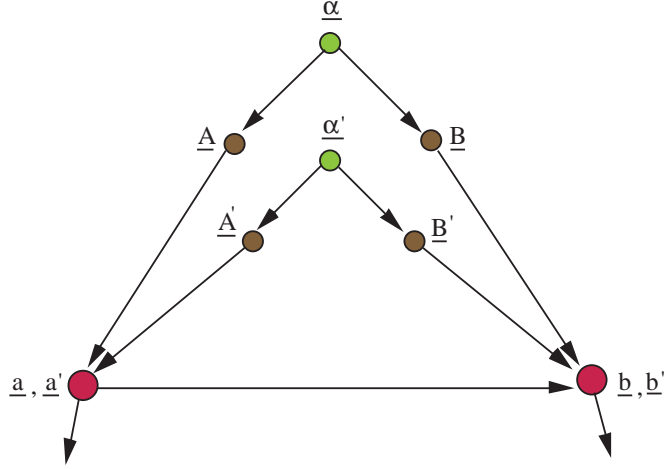, height=2.5in}
    \caption{Classical Bayesian Net that motivates
    the definition of $E_D$.}
    \label{fig:ent-dist}
    \end{center}
\end{figure}

Consider the Classical Bayesian Net of Fig.\ref{fig:ent-dist}.
The arrow from $(\rva, \rva')$ to $(\rvb, \rvb')$
allows what is often referred to as
``classical communication from Alice to Bob".
Let $\rvX =(\rvA, \rvB)$,
$\rvX' =(\rvA', \rvB')$.
Let
$N_\rvA=N_{\rvA'}=N_\rva=N_{\rva'}$ and
$N_\rvB=N_{\rvB'}=N_\rvb=N_{\rvb'}$.
The net of
Fig.\ref{fig:ent-dist} satisfies:
\beq
P(a, b, a', b') =
\sum_{X, X'}
P(a, a'| A, A')
P(b, b' |B, B', a, a')
P(X)
P(X')
\;,
\label{eq:prob-ent-dist}\eeq
where

\begin{subequations}
\label{eq:prob-sepa}
\beq
P(X) = \sum_\alpha P(A|\alpha)P(B|\alpha)P(\alpha)
\;,
\eeq
and

\beq
P(X') = \sum_{\alpha'} P(A'|{\alpha'})P(B'|{\alpha'})P({\alpha'})
\;.
\eeq
\end{subequations}
We wish to consider only those experiment in which
$\rva'$ and $\rvb'$
are both fixed at a known value, call it 0 for definiteness.
For such experiments, one considers:

\beq
P(a, b | a'=b'=0) =
\frac{P(a, b, a'=0, b'=0)}
{P(a'=0, b'=0)}
\;.
\eeq
Henceforth we will use $\Gamma$ as a short-hand for the
string ``$\rva'=0, \rvb'=0$". We will also use
$U$ to denote $P_{\rva,\rva'|\rvA,\rvA'}$ and
$V$ to denote $P_{\rvb,\rvb'|\rvB,\rvB', \rva, \rva'}$.

In the classical case, we define $E_D$ by

\beq
E_D(P_\rvX, P_{\rvX'})
=
\max_{U, V}
\min_{P_{\rva, \rvb, \rvlam | \Gamma}\in K}
H(\rva : \rvb | \rvlam, \Gamma)
\;,
\eeq
where $K$ is the set of all probability distributions
$P_{\rva, \rvb, \rvlam | \Gamma}$
with a fixed marginal $P_{\rva, \rvb| \Gamma}$ that
 satisfies Eq.(\ref{eq:prob-ent-dist}).
$P_{\rva, \rvb| \Gamma}$ depends on
$P_\rvX, P_{\rvX'}, U, V$. Since we maximize over $U,V$,
$E_D$ is a function of $P_\rvX$ and $P_{\rvX'}$.

Next we will show that the
 net of Fig.\ref{fig:ent-dist},
 {\bf without the classical communication arrow},
 satisfies:

\beq
E_D(P_\rvX, P_{\rvX'})\leq
E_F(P_\rvX) + E_F(P_{\rvX'})
\;.
\label{eq:ed-lt-ef}\eeq
Suppose we could show that

\beq
H(\rva: \rvb|\rvalp, \rvalp', \Gamma)\leq
H((\rvA, \rvA') : (\rvB,\rvB')| \rvalp, \rvalp')
\;.
\label{eq:cmis-goal}\eeq
After taking limits on the left hand side, this gives

\beq
E_D \leq
H(\rvA, \rvA' : \rvB,\rvB'| \rvalp, \rvalp')
\;.
\label{eq:cmis-left}\eeq
Note that by the independence of the prime and unprimed variables

\beq
H(\rvA, \rvA' : \rvB,\rvB'| \rvalp, \rvalp')
=
H(\rvA:\rvB|\rvalp)
+
H(\rvA':\rvB'|\rvalp')
\;.
\label{eq:cmis-right}\eeq
Eqs.(\ref{eq:cmis-left}) and (\ref{eq:cmis-right})
imply
Eq.(\ref{eq:ed-lt-ef}). So let us concentrate on establishing
Eq.(\ref{eq:cmis-goal}).
Events $\rvA, \rvA', \rvB,\rvB'$
all occur before $\Gamma$ so they are independent of $\Gamma$.
Therefore, we can write:

\beq
H(\rvA, \rvA' : \rvB,\rvB'| \rvalp, \rvalp')
=
H(\rvA, \rvA' : \rvB,\rvB'| \rvalp, \rvalp', \Gamma)
\;.
\label{eq:gamma-indep}\eeq
Because of Eq.(\ref{eq:gamma-indep}),
Eqs.(\ref{eq:cmis-goal}) is equivalent to:

\beq
H(\rva : \rvb| \rvalp, \rvalp', \Gamma)
\leq
H(\rvA, \rvA' : \rvB,\rvB'| \rvalp, \rvalp', \Gamma)
\;.
\label{eq:cmis-goal2}\eeq
Eq.(\ref{eq:cmis-goal2}) follows easily from the following Lemma,
which is proven in Appendix\ref{app:dpis}. Lemma:
The net of Fig.\ref{fig:data-pro}  satisfies

\beq
H(\rva: \rvb |\rvlam) \leq
H(\rvx: \rvy |\rvlam)
\;.
\label{eq:dat-pro-cmi}\eeq

\begin{figure}[h]
    \begin{center}
    \epsfig{file=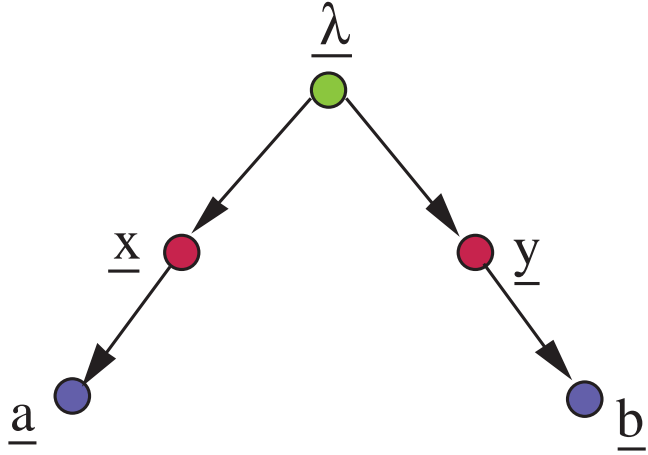, height=2.5in}
    \caption{Classical Bayesian Net that obeys
    the Data Processing Inequality Eq.(\ref{eq:dat-pro-cmi}).}
    \label{fig:data-pro}
    \end{center}
\end{figure}

\section{Quantum Distillation}
In this section, we will give a quantum counterpart of
the classical $E_D$ defined  in the previous section.

As in the classical case, let
$\rvX =(\rvA, \rvB)$,
$\rvX' =(\rvA', \rvB')$.
Let
$N_\rvA=N_{\rvA'}=N_\rva=N_{\rva'}$ and
$N_\rvB=N_{\rvB'}=N_\rvb=N_{\rvb'}$.
Suppose
$\rho_\rvX\in \dm(\hil_\rvX)$ and
$\rho_{\rvX'}\in \dm(\hil_{\rvX'})$ are given.
Suppose $U_{\rva, \rva'| \rvA, \rvA'}$ is a unitary
transformation mapping $\hil_{\rvA, \rvA'}$ onto $\hil_{\rva, \rva'}$:
\beq
U^{\dagger}_{\rvA, \rvA'| \rva, \rva'}
U_{\rva, \rva'| \rvA, \rvA'}
 = 1
\;.
\eeq
Likewise, suppose that for each $a$,
$V^a_{\rvb, \rvb'| \rvB, \rvB'}$is a unitary
transformation mapping $\hil_{\rvB, \rvB'}$ onto $\hil_{\rvb, \rvb'}$:
\beq
V^{\dagger a}_{\rvB, \rvB'| \rvb, \rvb'}
V_{\rvb, \rvb'| \rvB, \rvB'}^a
= 1
\;.
\eeq
Define the following projector on $\hil_\rva$:
\beq
\pi_\rva^a= \ket{a}_\rva \bra{a}_\rva
\;.
\eeq
Now consider the following density matrix

\beq
\rho_{\rva, \rvb| \Gamma} = \frac{1}{P(\Gamma)}
\sum_a
\bra{0_{\rva'}, 0_{\rvb'}}
\pi_\rva^a
U_{\rva, \rva'| \rvA, \rvA'}
V_{\rvb, \rvb'| \rvB, \rvB'}^a
\rho_\rvX
\rho_{\rvX'}
V^{\dagger a}_{\rvB, \rvB'| \rvb, \rvb'}
U^{\dagger}_{\rvA, \rvA'| \rva, \rva'}
\pi_\rva^a
\ket{0_{\rva}, 0_{\rvb'}}
\;,
\label{eq:rho1-ent-dist}\eeq
where $P(\Gamma)$ is defined so that
$\tr_{\rva,\rvb}(\rho_{\rva, \rvb| \Gamma}) =1$.
The previous equation can also be expressed in index notation as:

\beqa
\av{a, b_1| \rho_{\rva, \rvb|\Gamma}| a, b_2}
&=&
\frac{1}{P(\Gamma)}
\av{a, a'=0 | U_{\rva, \rva'| \rvA, \rvA'}| \suma{A_1}, \suma{A'_1}}\\
&&
\av{b_1, b'=0 | V^{a}_{\rvb, \rvb'| \rvB, \rvB'}| \suma{B_1}, \suma{B'_1}}\\
&&
\av{\suma{A_1}, \suma{B_1} |\rho_\rvX| \suma{A_2}, \suma{B_2}}\\
&&
\av{ \suma{A_1'}, \suma{B'_1}|\rho_{\rvX'}| \suma{A'_2}, \suma{B'_2}}\\
&&
\av{\suma{B_2}, \suma{B'_2}|
V^{\dagger a}_{\rvB, \rvB'| \rvb, \rvb'}
|b_2, b'=0}\\
&&
\av{\suma{A_2}, \suma{A'_2}|
U^{\dagger}_{\rvA, \rvA'| \rva, \rva'}
|a, a'=0}
\;.
\eeqa
Finally, we define $E_D$ by

\beq
E_D(
\rho_\rvX,
\rho_{\rvX'})
= \max_{U, V} \min_{\rho_{\rva, \rvb, \rvlam|\Gamma}\in K}
S(\rva: \rvb| \rvlam, \Gamma)
\;,
\label{eq:quan-ent-dist}\eeq
where $K$ contains all density matrices
$\rho_{\rva, \rvb, \rvlam|\Gamma}$ with a fixed marginal
$\rho_{\rva, \rvb|\Gamma}$ that
satisfies Eq.(\ref{eq:rho1-ent-dist}).

\begin{appendix}
\section{Appendix: Some Data Processing Inequalities}\label{app:dpis}
In this appendix, we will prove
two well known Data Processing Inequalities.

If $P, Q \in \pd(S_\rvx)$, then
the {\it relative entropy} (or Kullback-Leibler distance)
between $P$ and $Q$ is
\beq
D(P//Q) = \sum_x P(x) \ln [P(x)/Q(x)]
\;.
\eeq

\begin{lemma}\label{lemma:dpi-re}
(Data Processing Inequality for Relative Entropy, see Ref.\cite{dpi-rel-ent})
If $P, Q \in \pd(S_\rvx)$ and $T= \{T(y|x): y\in S_\rvy, x\in S_\rvx\}$
is a matrix of non-negative numbers such that $\sum_y T(y|x)=1$,
then
\beq
D(P//Q)\geq D(TP//TQ)
\;,
\eeq
where $TP$ should be understood as the matrix product of the column
vector $P$ times the matrix $T$.
\end{lemma}
\noindent proof:

One has
\beqa
D(P//Q) &=& \sum_x P(x) \ln \left(\frac{P(x)}{Q(x)}\right)\\
&=& \sum_y \sum_x T(y|x) P(x) \ln
\left(
\frac{T(y|x) P(x)}{T(y|x)Q(x)}
\right)\\
\label{line:log-sum}
&\geq& \sum_y TP(y) \ln
\left(
\frac{TP(y)}{TQ(y)}
\right) = D(TP//TQ)
\;.
\eeqa
Line (\ref{line:log-sum}) follows from an
application of the Log-Sum Inequality\cite{log-sum-ineq}.
QED

\begin{lemma}\label{lemma:dpi-cmi}
(Data Processing Inequality  for CMI)
The net of Fig.\ref{fig:data-pro}  satisfies

\beq
H(\rvx:\rvy|\rvlam) \geq H(\rva:\rvb|\rvlam)
\;.
\eeq
\end{lemma}
proof:

 The net of Fig.\ref{fig:data-pro} represent a
 probability distribution of the form:
\beq
P(a, b, x, y, \lambda) =
P(a|x)P(b|y)
P(x|\lambda)P(y|\lambda)
P(\lambda)
\;.
\eeq
One can easily show that such a probability distribution
satisfies:

\begin{subequations}
\beq
P(a|\lambda) = \sum_x P(a|x)P(x|\lambda)
\;,
\eeq

\beq
P(b|\lambda) = \sum_y P(b|y)P(y|\lambda)
\;,
\eeq
\label{eq:markov}
\end{subequations}

\begin{subequations}
\beq
P(a, b|\lambda) = P(a|\lambda) P(b|\lambda)
\;,
\eeq

\beq
P(x, y|\lambda) = P(x|\lambda) P(y|\lambda)
\;.
\eeq
\label{eq:indep}
\end{subequations}

For any two random variables $\rvlam, \rvx$,
let $Q^\lambda_\rvx$ be shorthand for $P_{\rvx|\rvlam=\lambda}$.
In other words,
$Q^\lambda_\rvx(x) = P_{\rvx|\rvlam}(x|\lambda)$ for all $x, \lambda$.
The two CMI we are dealing with  can be rewritten in terms of
relative entropy as follows:
\beq
H(\rvx: \rvy| \rvlam) = \sum_\lambda P(\lambda)
D( Q^\lambda_{\rvx,\rvy} // Q^\lambda_{\rvx}Q^\lambda_{\rvy})
\;,
\eeq
and

\beq
H(\rva: \rvb| \rvlam) = \sum_\lambda P(\lambda)
D( Q^\lambda_{\rva,\rvb} // Q^\lambda_{\rva}Q^\lambda_{\rvb})
\;.
\eeq
Thus, if we can show that
\beq
D( Q^\lambda_{\rvx,\rvy} // Q^\lambda_{\rvx}Q^\lambda_{\rvy})
\geq
D( Q^\lambda_{\rva,\rvb} // Q^\lambda_{\rva}Q^\lambda_{\rvb})
\;,
\eeq
then the present Lemma will be proven. The last inequality
will follow from Lemma \ref{lemma:dpi-re}
if we can find a transition probability matrix
$T(a,b|x,y)$ such that

\begin{subequations}
\beq
P(a, b|\lambda) =
\sum_{x,y}
T(a,b|x,y)
P(x, y|\lambda)
\;,
\eeq
and

\beq
P(a|\lambda) P(b|\lambda) =
\sum_{x,y}
T(a,b|x,y)
P(x|\lambda) P(y|\lambda)
\;.
\eeq
\label{eq:t-constraints}
\end{subequations}
Eqs.(\ref{eq:t-constraints}) follow easily
from Eqs.(\ref{eq:markov}) and (\ref{eq:indep}),
with $T$ given by :

\beq
T(a,b|x,y) = P(a|x) P(b|y)
\;.
\eeq

\noindent QED

\end{appendix}

\end{document}